\documentclass[%
 reprint,
superscriptaddress,
amsmath,amssymb,
aps, 
prb,
floatfix,
]{revtex4-2}

\usepackage{graphicx}
\usepackage{dcolumn}
\usepackage{bm}
\usepackage{hyperref}
\hypersetup{colorlinks, allcolors=blue}
\usepackage{physics}
\usepackage[version=4]{mhchem}
\usepackage{siunitx}
\usepackage[dvipsnames]{xcolor}
\usepackage{float}

\begin{document} 

\preprint{PRB}

\title{The diamond (111) surface reconstruction and epitaxial graphene interface}
\author{B.~P.~Reed}
 \affiliation{Department of Physics, Aberystwyth University, Aberystwyth, SY23 3BZ, United Kingdom.}
 \affiliation{Centre for Doctoral Training in Diamond Science and Technology, University of Warwick, Coventry, CV4 7AL, United Kingdom.}
 \affiliation{National Physical Laboratory, Teddington, TW11 0LW, United Kingdom.}
\author{M.~E.~Bathen}
 \affiliation{Advanced Power Semiconductor Laboratory, ETH Zurich, Physikstrasse 3, Zurich, 8092, Switzerland.}
 \affiliation{Centre for Materials Science and Nanotechnology, University of Oslo, Oslo, 0318, Norway.}
\author{J.~W.~R.~Ash}
 \affiliation{Department of Physics, Aberystwyth University, Aberystwyth, SY23 3BZ, United Kingdom.}
 \affiliation{Centre for Doctoral Training in Diamond Science and Technology, University of Warwick, Coventry, CV4 7AL, United Kingdom.}
\author{C.~J.~Meara}
  \affiliation{Centre for Doctoral Training in Diamond Science and Technology, University of Warwick, Coventry, CV4 7AL, United Kingdom.}
 \affiliation{School of Electrical and Electronic Engineering, Newcastle University, Newcastle-upon-Tyne, NE1 7RU, United Kingdom.}
\author{A.~A.~Zakharov}
 \affiliation{Max IV Laboratory, Lund University, Lund, 221 00, Sweden.}
\author{J.~P.~Goss}
 \affiliation{School of Electrical and Electronic Engineering, Newcastle University, Newcastle-upon-Tyne, NE1 7RU, United Kingdom.}
\author{J.~W.~Wells}
 \affiliation{Centre for Materials Science and Nanotechnology, University of Oslo, Oslo, 0318, Norway.}
\author{D.~A.~Evans}
 \affiliation{Department of Physics, Aberystwyth University, Aberystwyth, SY23 3BZ, United Kingdom.}
\author{S.~P.~Cooil}
\email[Corresponding author email:~]{scooil@icloud.com}
\altaffiliation[\\Current address:~] {Centre for Materials Science and Nanotechnology, University of Oslo, Oslo, 0318, Norway.}
 \affiliation{Department of Physics, Aberystwyth University, Aberystwyth, SY23 3BZ, United Kingdom.}
 \affiliation{Centre for Materials Science and Nanotechnology, University of Oslo, Oslo, 0318, Norway.} 
 
\date{\today}

\begin{abstract}
The evolution of the diamond (111) surface as it undergoes reconstruction and subsequent graphene formation is investigated with angle-resolved photoemission spectroscopy, low energy electron diffraction, and complementary density functional theory calculations. The process is examined starting at the C$(111)\text{-}(2\times1)$ surface reconstruction that occurs following detachment of the surface adatoms at \SI{920}{\degreeCelsius}, and continues through to the liberation of the reconstructed surface atoms into a free-standing monolayer of epitaxial graphene at temperatures above \SI{1000}{\degreeCelsius}. Our results show that the C$(111)\text{-}(2\times1)$ surface is metallic as it has electronic states that intersect the Fermi-level. This is in strong agreement with a symmetrically $\pi\text{-bonded}$ chain model and should contribute to resolving the controversies that exist in the literature surrounding the electronic nature of this surface. The graphene formed at higher temperatures exists above a newly formed C$(111)\text{-}(2\times1)$ surface and appears to have little substrate interaction as the Dirac-point is observed at the Fermi-level. Finally, we demonstrate that it is possible to hydrogen terminate the underlying diamond surface by means of plasma processing without removing the graphene layer, forming a graphene-semiconductor interface. This could have particular relevance for doping the graphene formed on the diamond (111) surface via tuneable substrate interactions as a result of changing the terminating species at the diamond-graphene interface by plasma processing.
\end{abstract}

\maketitle

\section{Introduction}
A rejuvenated interest in diamond as an electronic material has occurred over the last couple of decades as its importance for emerging state-of-the-art technologies continues to become apparent. For example, the nitrogen vacancy (NV) centres in diamond are one of the prime candidates for qubit storage in quantum computers \cite{Bradley2019}. Within the field of diamond electronics, the majority of studies have focused on the (100) orientation as it is the preferred growth direction for synthesized material by low pressure chemical vapour deposition (CVD) \cite{Ashfold1994} and, until recently, atomically flat surfaces have been easier to prepare \cite{Rawles1997}. However, the (111) surface has many desirable properties compared to the other low index surfaces: a higher Hall electron mobility in phosphor doped thin-films \cite{Pernot2006}, a higher temperature superconducting transition suggested to result from a non-uniform lattice expansion at high boron concentrations \cite{Takano2007,Okazaki2015}, and a higher sheet carrier density (hole mobility) for the hole accumulation layer on hydrogen-terminated surfaces \cite{Hirama2010}; Finally, the average alignment of NV centres can be controlled during CVD growth on (111) substrates to a single crystallographic direction, increasing the homogeneity of the NV centre and maximizing the collection efficiency of NV emission in bulk single crystal samples. This is ideal for quantum information and sensing applications, naturally securing (111) orientated diamond’s future in the development of quantum devices \cite{Michl2014,Lesik2014}.

Despite these attractive attributes, fundamental studies of the diamond (111) surface remain incomplete. For example, with respect to the reconstructed C$(111)\text{-}(2\times1)$ surface, controversy surrounding the electronic nature of the surface has arisen by trying to match the published experimental angle-resolved photoemission spectroscopy (ARPES) results to density functional theory (DFT) calculations. In order to accomodate the semiconducting nature of the surface states observed experimentally \cite{Kern1998,Graupner1997,Himpsel1980,Himpsel1981}, structural models suggesting dimerized or buckled $\pi\text{-bonded}$ surface chains have been presented \cite{Pandey1982}. The resulting electronic structure from such models exhibits a pronounced bandgap opening in the surface state that is well matched to the experimental results. However, the most energetically favourable structural models suggest that symmetric $\pi\text{-bonded}$ surface chains with little-to-no dimerization should occur \cite{Bechstedt2001,Schmidt1996,Vanderbilt1984}, and that the resulting electronic structure should feature a metallic surface state that intersects the Fermi-level ($E_{\text{F}}$). The most recent measurements of the C$(111)\text{-}(2\times1)$ surface structure using low energy electron diffraction (LEED) do suggest a lack of dimerization, indicating that a surface with symmetric $\pi\text{-bonded}$ chains is accurate \cite{Walter2002}. However, measurements of the electronic structure that exhibit the expected metallic surface state indicative of this structural model have not been presented to date. In this article we aim to provide clear evidence of the metallic nature of the surface states by use of synchrotron-based ARPES studies compared to complementary DFT calculations, along with LEED and core-level X-ray photoelectron spectroscopy (XPS) measurements. 

The growing demand for miniaturisation has also led to increased interest in the nanoscale properties of diamond \cite{PakpourTabrizi2020} and the production of diamond-graphene ($\text{sp}^3$~--~$\text{sp}^2$) interfaces, investigated both experimentally \cite{Wan2020,Ogawa2012,Norio2013,Wu2011,Shiga2012,Yu2012,Song2018,Jungnickel1996,Wang2012,Selli2013} and through DFT studies \cite{Zhao2016,Zhao2019,Ma2012,Hu2013}. There now exists a multitude of graphene growth systems with the ability to produce high quality epitaxial graphene. Of relevance to this work are those methods for which the supply of carbon lies within the substrate. These primarily include high temperature treatment of SiC and the segregation of carbon from bulk single crystal metals (e.g. Ni, Ru, Ir, and Pt). Direct measurements of the electronic structure of graphene prepared by each method was necessary to assess its suitability for future electronic applications. For example, the quality and electronic properties of graphene grown on SiC is highly dependent on the chosen growth face \cite{Luxmi2010,Varchon2007} whilst doping \cite{Martinez2016}, the opening of a band-gap at the Dirac-point \cite{Zhou2007}, and mini-gaps elsewhere in the electronic structure \cite{Pletikosic2009} have been shown to arise from a variety of graphene substrate interactions. However for graphene grown directly on the diamond (111) surface, we find no direct measurement of the electronic structure in the literature. To this end, we will proceed to follow the C$(111)\text{-}(2\times1)$ system through to the detachment of the reconstructed surface atoms, and their promotion to a freestanding graphene layer in order to investigate the resulting electronic structure.

\begin{figure*}[t]
 \centering
 \includegraphics[width=\textwidth]{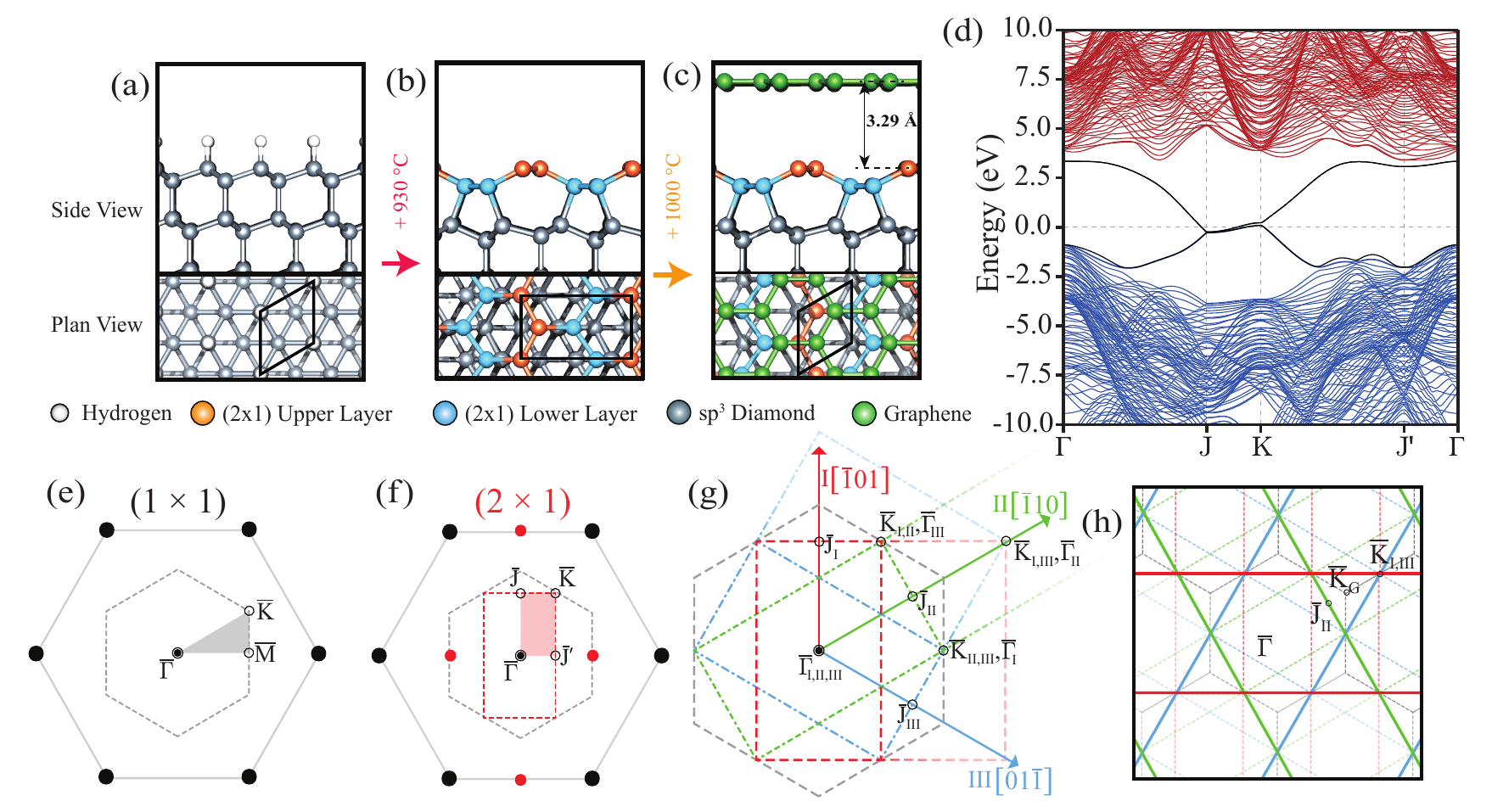}
 \caption{Structure models, surface quality, and schematics detailing the nomenclature of the BZs used throughout the article. (a)--(c) The optimized lowest energy configuration positions calculated by DFT. Namely, the C$(1\times1)\text{-}(111)$:H surface, the C$(111)\text{-}(2\times1)$ symmetric $\pi\text{-bonded}$ chain reconstructed surface, and the graphitized C$(111)\text{-}(2\times1)$ surface respectively. The surface primitive unit cells are indicated by the black solid lines on each plan view. (d) DFT calculated electronic structure of the symmetrically $\pi\text{-bonded}$ C$(111)\text{-}(2\times1)$ surface, with the surface state shown in black. (e)--(f) Schematics of the reciprocal space patterns for the $(1\times1)$ and $(2\times1)$ lattice respectively, along with the conventional notation used for the high symmetry points of the BZs. The black circles represent the reciprocal lattice originating from the diamond $(1\times1)$ surface and the grey dashed hexagon is the surface BZ with its irreducible part shaded in grey. In (f), the red coloured circles represent the new $(2\times1)$ surface reciprocal lattice that arises following reconstruction and the red dashed rectangle is the new surface BZ with its irreducible part shaded in red. (g) Surface BZ for the three-domain C$(111)\text{-}(2\times1)$ surface along with the notation used when high symmetry points overlap as a result of the spatially averaging nature of the photoemission techniques used. Each colour (red, blue, and green) represents one orientation of the $(2\times1)$ BZ. (h) Multiple adjacent BZs with solid-coloured lines that represent the almost flat band that traverses the short edge of the rectangular $(2\times1)$ zone in red, green, and blue for their respective rotations. Averaging over space would lead to a repeating hexagram represented by all colours, at binding energies close to $E_{\text{F}}$. The location of the high symmetry points at the corner of graphene’s hexagonal BZ, $\overline{\text{K}}_{\text{G}}$, is also shown relative to the $(2\times1)$ BZs.} \label{fig:struct-mod}
\end{figure*}

\section{Experiment and Calculations}
\subsection{Experimental Details}
Two single crystal diamond (111) samples with a low miscut angle to the (111) plane were prepared by mechanical polishing by Element Six (Harwell, UK). The miscut angle was found to be $\pm$\ang{0.5} by Laue measurements. The root-mean-square roughness of the diamond surfaces, determined by contact mode atomic force microscopy, was found to be \SI{0.16}{\nano\metre} and \SI{0.20}{\nano\metre} along and perpendicular to the polishing lines respectively. A natural type IIb sample with a boron concentration of \SI{e15}{\per\centi\metre\cubed} 
was used for the ARPES and ultra-violet photoelectron spectroscopy (UPS) measurements, and a synthetic high-pressure high-temperature (HPHT) type Ia sample was used  
for XPS measurements. Samples were initially hydrogen terminated in a custom low-power (\SI{150}{\watt}) capillary discharge plasma system with open geometry. H-termination was confirmed by the measurement of a sharp C$(1\times1)$ LEED pattern, \cite{Pate1986} as well as an increase in the secondary electron tail measured by UPS, which is characteristic of a negative electron affinity (NEA) of this surface. \cite{Bandis1995}

ARPES measurements were performed at beamline SGM3 of the ASTRID synchrotron radiation facility (Aarhus, Denmark).
Typically for ARPES measurements, low photon energies $<$ \SI{50}{\electronvolt} are preferred as this provides a higher photoionization cross section and an improved energy and momentum resolution \cite{Andrea2004}. However, diamond is a peculiar case where there are a lack of free electron like final states (FEFS) available when using low excitation energies, therefore necessitating the use of higher photon energies \cite{Himpsel1980}. Several unsuitable energy ranges have been detailed \cite{PakpourTabrizi2020}, particularly the use of photon energies typically available in home laboratories, such as the He(I) and He(II) emission lines at \SI{21.2}{\electronvolt} and \SI{40.08}{\electronvolt} respectively. Our ARPES measurements are recorded from a plane close to the centre of a 3D Brillouin zone (BZ) for ease of comparison to the DFT calculations. To achieve this, measurements of the valence band maximum (VBM) were recorded whilst varying the photon energy until the bulk $\sigma\text{-bands}$ had their maximum closest to $E{_\text{F}}$, corresponding to $h\nu=$ \SI{125}{\electronvolt}. Binding energies are referenced relative to $E{_\text{F}}$ measured on an \ce{Ar+} sputter-cleaned Cu crystal.

\begin{figure*}[t]
 \centering
 \includegraphics[width=\textwidth]{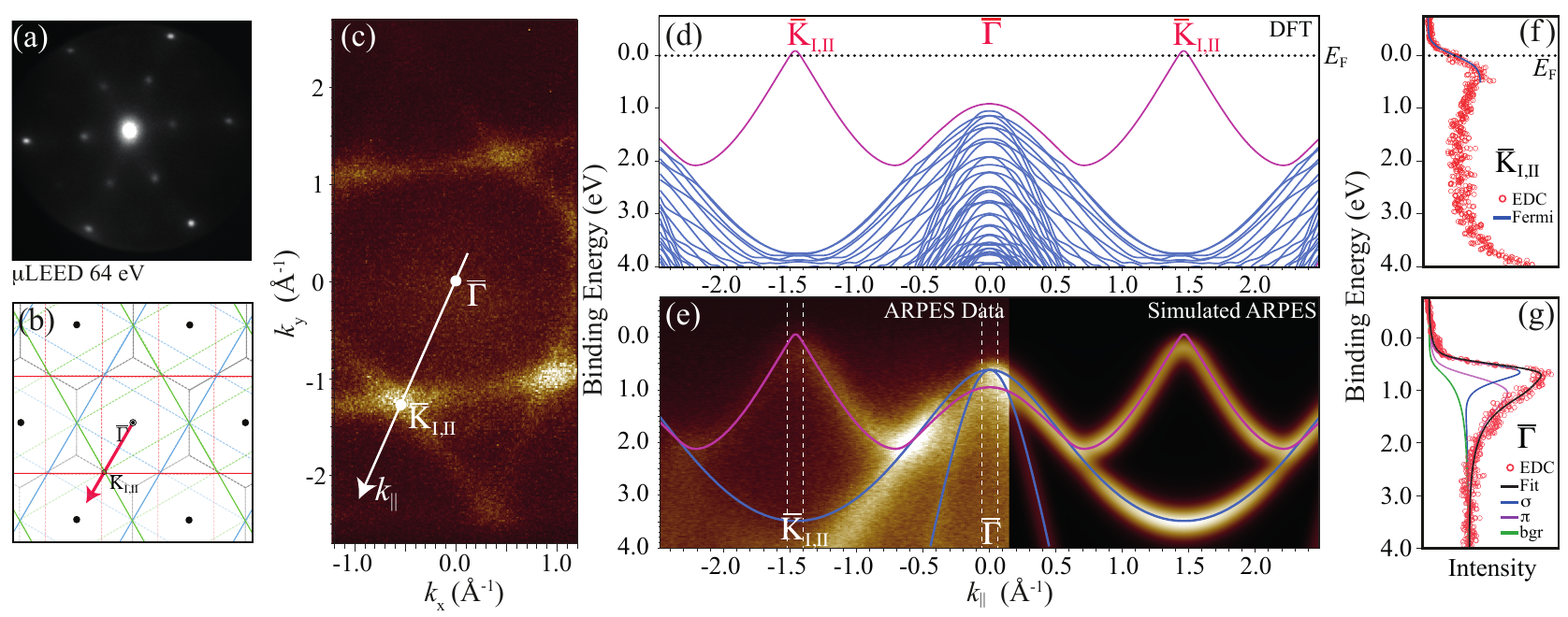}
 \caption{$\mu$LEED, ARPES, and DFT results from the C$(111)\text{-}(2\times1)$ surface reconstruction following in vacuo annealing at \SI{920}{\degreeCelsius}. (a) $\mu$LEED pattern acquired at a kinetic energy of \SI{64}{\electronvolt}, spots from all three rotational domains are observed with comparable intensity. (b) Schematic diagram of the $\overline{\Gamma}\text{-}\overline{\text{K}}_{\text{I,II}}$ direction through the surface BZs used for the DFT and ARPES datasets presented in panels (d) and (e), respectively. (c) Constant energy surface from the ARPES dataset at $E_{\text{B}}$ = \SI{0.1}{\electronvolt}. The white arrow indicates the direction of the slice used to produce the $E$ vs $k_{\parallel}$ data shown in (e). (d) DFT results of the occupied valence band structure, the $\pi\text{-band}$ originating from the reconstruction is shown in magenta and the bulk $\sigma\text{-bands}$ in blue. (e) Photoemission intensity on the left-hand side and the simulated intensity on the right. Overlaid on the image is the $\pi\text{-band}$ that results from the 64 atom DFT supercell calculation in magenta, along with the bulk $\sigma\text{-bands}$ from the simple 2-atom unit cell DFT calculation in blue. The white vertical lines are the boundaries of the integrated area used for the EDCs at $\overline {\text{K}}_{\text{I,II}}$ and $\overline{\Gamma}$ shown in panels (f) and (g), respectively.} \label{fig:2by1-surf}
\end{figure*}

$\mu$LEED measurements were performed at beamline I311 at the MaxLab synchrotron (Lund, Sweden) using an Elmitec III SPELEEM photoemission and low energy electron microscope. Atomic hydrogen was provided by a Tectra atomic hydrogen source within the UHV system at a pressure of \SI{5e-9}{\milli\bar}. XPS and UPS measurements were carried out in the Aberystwyth University laboratory's real-time electron emission spectroscopy (REES) system equipped with a Specs PHOIBOS 100 electron analyzer, 2D CCD detector, and non-monochromated dual anode X-ray source. The temperatures reported from this system are measured by a type-K thermocouple, for which the readings have been calibrated by means of in-situ Raman spectroscopy measurements as detailed in references \cite{Liu2000,Cui1998}, see supplementary information Fig.~S1 for more details.

\subsection{Computational Details}
The electron dispersions, total cohesion energies, and geometrical structures were calculated using DFT employing the generalised gradient approximation (GGA) implemented in the Perdew-Burke-Ernzerhof (PBE) exchange correlation functional \cite{Perdew1996}. This was done using the {\sc{Aimpro}} \cite{Briddon2011} and {\sc{Vasp}} \cite{Kresse1996} packages to ascertain if any differences manifest when including Van der Waals (VdW) interactions, as this has been suggested to be necessary for the graphene on diamond system \cite{Zhao2016, Zhao2019,Hu2013}.

For both codes, the surface was modeled by a supercell consisting of a slab with two surfaces and a vacuum region of \SI{30}{\angstrom}. The diamond slab was comprised of an orthorhombic unit cell of 60 $\text{sp}^3\text{-bonded}$ carbon atoms, with the top and bottom faces terminated with 4 $\text{sp}^2\text{-bonded}$ C$(111)\text{-}(2\times1)$ symmetric $\pi\text{-bonded}$ chain atoms. A graphene monolayer was then placed on either side of the slab by the addition of $8~\text{sp}^2\text{-bonded}$ carbon atoms. This structure was repeated ad infinitum in the $xy$-plane, and alternating layers of diamond and vacuum were then produced via repetition in the surface normal direction. The slab thickness of 32 atomic layers proved to be thick enough to prevent any electric field effects between opposite sides of the slab. The slab was carefully optimized allowing the atoms to adjust their positions until the maximum force on any atom was less than \SI{0.05}{\electronvolt\per\angstrom}. The electronic structures were sampled using a Monkhorst-Pack scheme \cite{Monkhorst1976} with a $8\times4\times1$ sampling grid. In the slab calculations performed, the number of bulk bands observed depends on the slab thickness, and would eventually approach a continuum for a very thick slab. \cite{Sagisaka2017}. The photoemission measurements however probe a small value of $k_\perp$ resulting in far fewer bands being observed in the measurement. In order to better compare the bulk bands of the DFT calculations and our measurements, we have also calculated the bulk band structure using a simple 2-atom unit cell repeated ad infinitum in $x$,$y$ and $z$. The bare $\sigma\text{-bands}$ were then extracted at $k_\perp =$ \SI{0}{\electronvolt\per\angstrom} and result in two bare bands representing the heavy and light hole bands. Throughout this paper, when a comparison between the DFT and the ARPES measurements is required, a constant Lorentzian broadening in momentum space has been applied to the DFT-calculated bare bands. This is achieved by use of a spectral function in which electron-defect interactions that limit the mean-free path of the carriers are assumed to be the dominant broadening mechanism. The comparatively low temperatures used during our ARPES measurements relative to diamond's much higher phonon temperature $(T_{\text{sample}} \ll T_{\text{phonon}})$ means that other quasiparticle interactions such as electron-phonon coupling, or electron-electron scattering are not assumed to strongly contribute to the measured line width. This broadening is achieved by setting the imaginary part of the self-energy $2\Sigma^{''} \cong \Gamma_{\text{e-df}}$ to a constant, which in turn, via the Kramers-Kronig relation requires the real part of the self energy term $\Sigma^{'}$ to have no particular structure. Here, $\Gamma_{\text{e-df}}$ is the inverse lifetime of the core-hole due to electron-defect scattering. The spectral function broadened bare bands have then been convolved by our experimental resolutions and multiplied by a normalized Fermi-function cut off at $E_{\text{B}} =$ \SI{0}{\electronvolt}, and a width appropriate for the $\sim$\SI{300}{K} measurements.

\section{Results and discussion}
The structural models relating to the evolution of the three thermodynamically stable stages of the diamond (111) surface whilst heating up to \SI{1000}{\degreeCelsius} are shown in Fig.~\ref{fig:struct-mod}(a)--(c). These models were constructed following optimisation of the atomic positions obtained from DFT calculations. The monohydride terminated C$(111)\text{-}(1\times1)$:H surface is shown in Fig:~\ref{fig:struct-mod}(a) and is the starting point for our experiments. The first double layer of the $(1\times1)$ surface is a corrugated hexagonal lattice with alternating surface and subsurface atoms, giving a rhombohedral primitive unit cell with lattice vector $a_0 =$ \SI{2.52}{\angstrom} as shown in Fig:~\ref{fig:struct-mod}(a). The resulting reciprocal space lattice is hexagonal as shown Fig:~\ref{fig:struct-mod}(e) along with the surface $(1\times1)$ BZ for which the calculated reciprocal space distances for $\overline{\Gamma}\text{-}\overline{\text{K}}$ and $\overline{\Gamma}\text{-}\overline{\text{M}}$ are \SI{1.66}{\per\angstrom} and \SI{1.44}{\per\angstrom} respectively. 

The desorption of H adatoms from the diamond’s surface results in unstable dangling bonds that cause a complex surface reconstruction as initially suggested by Pandey for the Si(111) surface \cite{Pandey1981}. The surface atoms from the upper and lower layer in the corrugated surface form symmetrically $\pi\text{-bonded}$ chains along the surface as shown in Fig.~\ref{fig:struct-mod}(b). This results in a real-space doubling of the surface periodicity in one surface mesh direction whilst maintaining bulk periodicity in the other, leading to a new rectangular $(2\times1)$ surface unit cell and BZ shown in Fig:~\ref{fig:struct-mod}(b) and (f). The calculated reciprocal distances of the high symmetry points from the zone center of the new $(2\times1)$ BZ are $\overline{\Gamma}\text{-}\overline{\text{K}}=$~\SI{1.44}{\per\angstrom}, $\overline{\Gamma}\text{-}\overline{\text{J}}=$~\SI{1.25}{\per\angstrom} and $\overline{\Gamma}\text{-}\overline{\text{J'}}=$~\SI{0.72}{\per\angstrom}. As a result of the three-fold symmetry of the hexagonal (111) surface, three rotational domains of the $(2\times1)$ reconstruction are possible, and due to the spatially averaging nature of the photoemission technique, all three will be captured concurrently. As in the previous experimental reports on this surface \cite{Graupner1997}, we will adopt a subscript of roman numerals when discussing the high symmetry points of the three $(2\times1)$ BZs, i.e. I, II, and III for the $[\bar{1}01]$, $[\bar{1}10]$, and $[01\bar{1}]$ directions shown in red, green, and blue respectively in Fig:~\ref{fig:struct-mod}(g). The labeling of the high symmetry points for the $(1\times 1)$ BZ of diamond will also be dropped as the only states close to $E_{\text{F}}$ are at the zone center.

Finally, at temperatures above \SI{1000}{\degreeCelsius} the surface reconstruction detaches forming a graphene layer above the diamond surface. This graphene layer exists above a newly formed C$(111)\text{-}(2\times1)$ as shown in Fig.~\ref{fig:struct-mod}(c) and results in a new hexagonal $(1\times1)$ surface BZ of graphene. In order to differentiate between the notation of the $(2\times1)$ surface BZ and that of the graphene, a subscript ‘G’ will be adopted when discussing its high symmetry points, with the exception of $\overline{\Gamma}$ as the centre of all the BZs discussed are congruent and will therefore not be presented with any subscript.

\subsection{The C$(111)\text{-}(2\times1)$ surface reconstruction}
The $\mu$LEED pattern shown in Fig.~\ref{fig:2by1-surf}(a) was acquired after annealing the diamond (111) surface to \SI{920}{\degreeCelsius} and is as expected when averaging three rotational domains of the C$(111)\text{-}(2\times1)$ reconstruction. ARPES measurements performed after the same temperature treatment, along with electron dispersions extracted from our DFT calculations for the C$(111)\text{-}(2\times1)$ surface, are presented in Fig.~\ref{fig:2by1-surf}. Pandey’s original symmetrically $\pi\text{-bonded}$ chain calculations \cite{Pandey1982} and ours shown in Fig.~\ref{fig:struct-mod}(d) indicate that the surface state is weakly dispersing with energy (almost flat), along the short edge of the rectangular $(2\times1)$ BZ , i.e. along $\overline{\text{J}}_{\text{I}}\text{-}\overline{\text{K}}_{\text{I}}$ in Fig.~\ref{fig:struct-mod}(f). This almost flat band should give rise to a constant energy surface at energies close to $E_{\text{F}}$ represented by a single colour in Fig.~\ref{fig:struct-mod}(h) e.g. blue, green or red. Whilst averaging all rotational domains will lead to a repeating hexagram pattern represented by all these colours. The hexagram constant energy surface is reproduced experimentally in Fig.~\ref{fig:2by1-surf}(c). The white arrow along $\overline{\Gamma}\text{-}\overline{\text{K}}_{\text{I,II}}$ indicates the direction through momentum space used for extracting the $E$ vs $k_{\parallel}$ data in Fig.~\ref{fig:2by1-surf}(d) and (e) from the DFT calculations and ARPES dataset respectively. The direction is also shown schematically in Fig.~\ref{fig:2by1-surf}(b), and reaches an equivalent $\overline {\text{K}}_{\text{I,II}}$-point as discussed in the work by Graupner \emph{et al.} \cite{Graupner1997}.

The DFT calculations presented in Fig.~\ref{fig:2by1-surf}(d) agree well with earlier modelling studies of the C$(111)\text{-}(2\times1)$ surface by several groups that predict symmetrically $\pi\text{-bonded}$ surface chains with states that intersect $E_{\text{F}}$ \cite{Kern1998,Bechstedt2001,Schmidt1996,Vanderbilt1984,Kern1996,Marsili2005,Iarlori1992}. For comparison with our collected ARPES data, the $\pi\text{-band}$ shown as a solid magenta line in Fig.~\ref{fig:2by1-surf}(d) is directly overlaid on the ARPES dataset shown in Fig.~\ref{fig:2by1-surf}(e), whereas the solid blue lines of the $\sigma\text{-bands}$ overlaid on the ARPES data are taken from the simple 2-atom bulk band calculation. It was necessary to perform a rigid energy shift of the $\sigma\text{-bands}$ so that their maxima lay \SI{0.66}{\electronvolt} below $E_{\text{F}}$, matching the position of the VBM of the C$(111)\text{-}(1\times1)$:H surface extracted from an energy distribution curve (EDC) at $\overline {\Gamma}$ (see supplementary Fig.~S2). These three bands have then been employed to generate the simulated ARPES dataset presented on the right of Fig.~\ref{fig:2by1-surf}(e). 

There is a strong agreement between the experimental and theoretical data, with the most obvious difference being the binding energy of the heavy hole $\sigma\text{-band}$ minima at $\overline {\text{K}}_{\text{I,II}}$. However, this is likely a result of underestimating the electron-electron correlation in the DFT calculations performed due to the self-interaction error of the PBE functional. EDCs taken at $\overline {\Gamma}$ and $\overline{\text{K}}_{\text{I,II}}$, averaged from a region \SI{0.1}{\per\angstrom} wide, are shown in Fig.~\ref{fig:2by1-surf}(f) and (g) respectively. The EDC of the band maxima at $\overline {\Gamma}$ is fitted with two asymmetric-Voigt profiles separated by \SI{0.33}{\electronvolt}, which is very similar to the energy separation found between the maxima of the $\pi$ and $\sigma$ bare-bands at $\overline {\Gamma}$ after compensating for the boron doping level in the sample by performing the rigid energy shift of the $\sigma\text{-bands}$ from our bulk diamond calculation. The asymmetry parameter is equal for both peaks however the FWHM of the $\pi\text{-band}$ is almost double that of the $\sigma\text{-band}$. An increase of the FWHM like this could result from an increased density of defects present within the Pandey chains. These defects most likely result from polishing lines or step edges and would strongly enhance the broadening caused by impurity scattering for the $\pi$ states. This is also not surprising given that the bulk states originate from a crystal of exceptional quality whilst the surface has been subjected to mechanical polishing. The EDC taken at $\overline{\text{K}}_{\text{I,II}}$ shows that the state is indeed metallic as its leading edge is best fit with a Fermi-function at \SI{0}{\electronvolt} and a width representative of a sample temperature of \SI{300}{K}. 

On the other hand, our ARPES measurements do not compare well with the experimental results presented by Graupner \emph{et~al.} and Himpsel \emph{et~al.} \cite{Graupner1997,Himpsel1981}. In their works, an energy gap of at least \SI{0.5}{\electronvolt} between the maximum of the surface state at $\overline{\text{K}}_{\text{I,II}}$ and $E_{\text{F}}$ is observed that led to the promotion of a dimerized-chain model by Pandey \cite{Pandey1982}. We can only speculative suggestions as to why these measurements show surface states with such an energy gap. For example, a wider energy gap could result from a larger miscut angle of their sample, therefore producing narrower (111) terraces and more step edges. Rougher diamond surfaces could cause buckling of the surface reconstructions, or a misalignment of the sample during the photoemission studies could be responsible. This does not however appear to be the case as, by taking cuts through our data sets at various rotations around $\overline {\Gamma}$, a gap of \SI{0.5}{\electronvolt} between the surface state and $E_{\text{F}}$ could not be reproduced. This is likely due to the almost flat nature of the states along the $\overline{\text{J}}\text{-}\overline{\text{K}}_{\text{I}}$ direction of the $(2\times1)$ rectangular surface BZ. 

\begin{figure}[p]
\centering
 \includegraphics[width=\columnwidth]{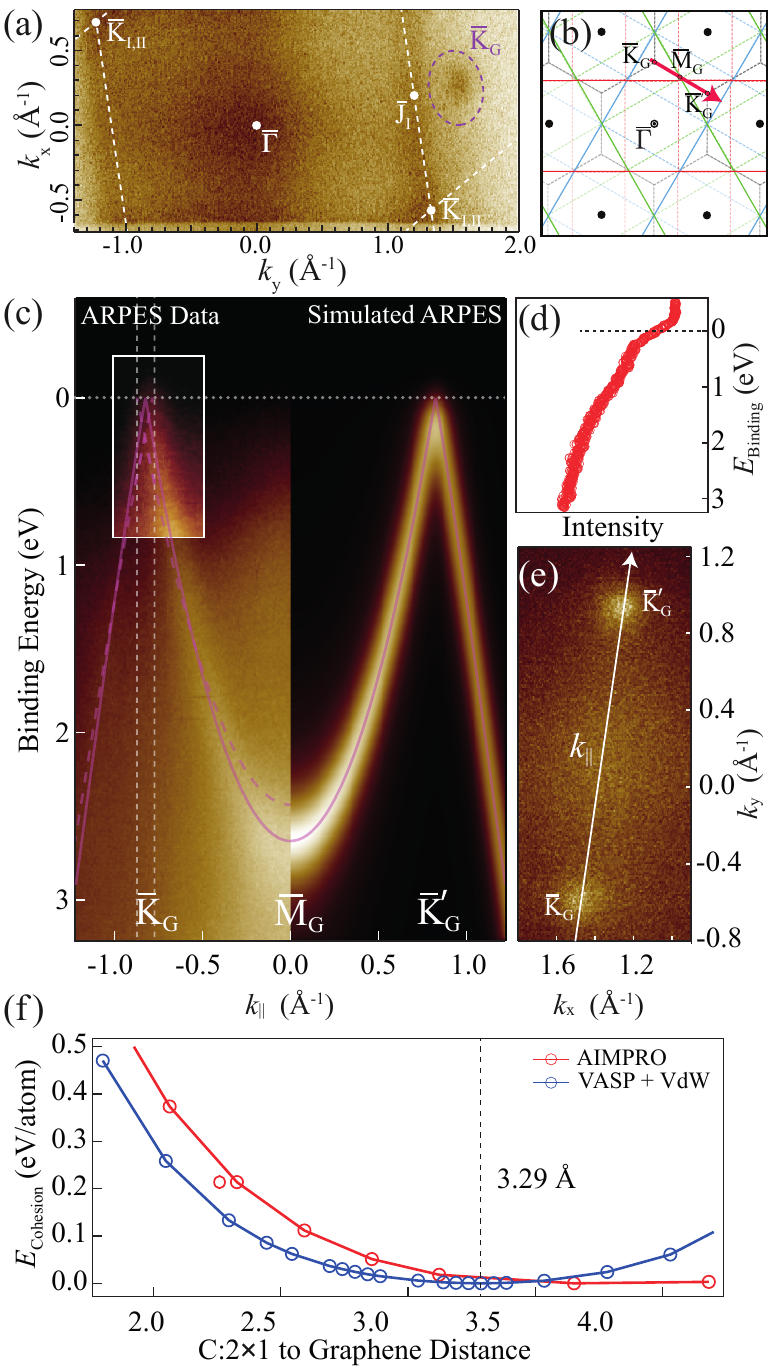}
 \caption{ARPES measurements and DFT calculations of graphene formed above the C:2$\times1$(111) surface. (a) Constant energy surface at $E_{\text{B}}=$~\SI{0.1}{\electronvolt} using $h\nu =$~\SI{125}{\electronvolt}. The graphene $\overline {\text{K}}_{\text{G}}$-point is circled in a magenta dotted line, and the white dotted lines are to guide the eye towards the hexagram features already discussed for the C$(111)\text{-}(2\times1)$ surface. (b) Schematic diagram showing the $\overline{\text{K}}_{\text{G}}\text{-}\overline{\text{M}}_{\text{G}} \text{-}\overline{\text{K}}\textsubscript{\rlap{G}}'~\text{-}\overline {\Gamma}\rlap{$'$}$ direction along the edge of graphene’s hexagonal BZ. (c) Photoemission intensity, DFT calculated bare bands and simulated intensity. The ARPES dataset on the left hand side was acquired with $h\nu =$~\SI{40.8}{\electronvolt}. Overlaid on the image are the DFT calculated $\pi\text{-band}$ as the dashed magenta line, and the same band after performing a rigid shift and stretch as a solid magenta line. The solid line is used for creating the simulated ARPES intensity on the right. (d) EDC taken at the $\overline {\text{K}}_{\text{G}}$ point with a width of \SI{0.1}{\per\angstrom} integrated between the vertical dashed white lines in (c). (e) Constant energy surface at $E_{\text{F}}$, using $h\nu =$~\SI{40.8}{\electronvolt}, the white arrow indicates the slice taken to extract the $E$~vs~$k_{\parallel}$ datasets shown in (c). (f) The graphene atoms' cohesion energy as a function of distance from the C$(111)\text{-}(2\times1)$ surface both with and without VdW interactions.} \label{fig:gra-ARPES}
\end{figure}

\subsection{Epitaxial graphene formation above the C$(111)\text{-}(2\times1)$ surface reconstruction}
At temperatures greater than \SI{1000}{\degreeCelsius}, the surface reconstructed atoms are liberated into a monolayer of freestanding graphene as shown in Fig.~\ref{fig:struct-mod}(c). New electron states coexisting with those already presented for the C$(111)\text{-}(2\times1)$ surface are observed in the ARPES datasets shown in Fig.~\ref{fig:gra-ARPES}. One of these features is shown in the constant energy surface shown in Fig.~\ref{fig:gra-ARPES}(a), located close to the corner of the hexagonal diamond ($1\times1$) BZ, where no states from the diamond C$(111)\text{-}(2\times1)$ electronic structure are expected. For increasing binding energy, the state displays the characteristic isoenergetic triangular modulation of the photoemission intensity, becoming more pronounced at energies away from $E_{\text{F}}$ \cite{Kruczy2008} as shown in Fig. S3(a), providing evidence that it originates from a graphene film. The feature labeled $\overline {\text{K}}_\text{G}$ appears at a distance of $k_{\parallel}=$~\SI{1.61}{\per\angstrom} from $\overline {\Gamma}$ which is close to the $\overline{\Gamma}\text{-}\overline{\text{K}}$ distance of the hexagonal diamond ($1\times1$) zone and in good agreement with the $\overline{\Gamma}\text{-}\overline{\text{K}}_{\text{G}}$ distance of freestanding graphene at $\SI{1.70}{\per\angstrom}$ \cite{Neto2009}. This is a good indication that the graphene is commensurate with the diamond surface. To decouple the graphene features from that of the diamond surface states, lower photon energies were used, making use of diamond's tendency to have weaker photoemission signal at low photon energies. The photoemission intensity along the $\overline{\text{K}}_{\text{G}}\text{-}\overline{ \text{M}}_{\text{G}}\text{-}\overline{\text{K}}\textsubscript{\rlap{G}}'$~ edge of the graphene BZ, (shown schematically in Fig.~\ref{fig:gra-ARPES}(b)), is presented in the left panel of Fig.~\ref{fig:gra-ARPES}(c). 

Using the structural model presented in Fig.~\ref{fig:struct-mod}(c) the electronic dispersion for the graphene on diamond system was calculated and found to be in good agreement with our experimental data, shown as the dashed magenta line overlaid on the ARPES dataset in Fig.~\ref{fig:gra-ARPES}(c). The main difference between the calculated electron dispersion and the experimental results appears to be the energy at which the band maximum and minimum are observed at the $\overline {\text{K}}_{\text{G}}$ and $\overline {\text{M}}_{\text{G}}$ points respectively. For the calculation to better match our experimental data, it was necessary to perform a linear stretching of the calculated energy axis by a factor of 1.2 and shift the energy of the band maximum at $\overline {\text{K}}_{\text{G}}$ from \SI{0.22}{\electronvolt} to $E_{\text{F}}$. The need to stretch the energy axis can be understood in terms of the commonly observed underestimation of electronic energies for GGA level exchange correlation functionals. This is shown as the solid magenta line in Fig.~\ref{fig:gra-ARPES}(c) and is the bare band used to simulate the ARPES data on the right-hand side of Fig.~\ref{fig:gra-ARPES}(c). The direction of the cut taken through the ARPES data unfortunately cuts through areas of weak photoemission intensity as a result of the triagonal warping mentioned earlier. However, the adjusted bare-band calculation matches well with the measured intensity as band progresses towards the next zone centre $\overline {\Gamma}\rlap{$'$}$, as shown in second differential image Fig.~S3(b). The graphene $\overline{\text{K}}_{\text{G}}$-points are observed at $k_{\parallel} =$~\SI{0.81}{\per\angstrom} (where $k_{\parallel} =$~\SI{0}{\per\angstrom} is reference to the saddle point of the band at the $\overline{\text{M}}_{\text{G}}$-point), giving a $\overline{\text{K}}_{\text{G}}\text{-}\overline{\text{K}}\textsubscript{\rlap{G}}'$ distance of ~\SI{1.62}{\per\angstrom} which is slightly smaller than that expected for freestanding graphene at \SI{1.70}{\per\angstrom} but within the accuracy of the $k$-warping performed in the analysis due to large emission angles. The continued presence of the $(2\times1)$ electronic states along with the features now present from the graphene provides strong evidence that epitaxial graphene is formed on the diamond (111) surface in co existence with the C$(111)\text{-}(2\times1)$ surface reconstruction. This high degree of epitaxial registry is further confirmed by the lack of any states related to rotational domains in our photoemission datasets. The constant energy surface taken at $E_{\text{F}}$ shown in Fig.~\ref{fig:gra-ARPES}(e) shows only two $\overline {\text{K}}_{\text{G}}$-points along the edge of graphene’s BZ and no other points that would indicate rotational domains seen on some metal \cite{Loginova2009,Suele2014} and semiconducting substrates \cite{Emtsev2008,First2010}. However, small scale rotational disorder such as that observed for epitaxial graphene on SiC(0001) \cite{Walter2013} cannot be ruled out. 

The $\overline {\text{K}}_{\text{G}}$-point intersects the Fermi level in the ARPES data, as seen in the high contrast inset in Fig.~\ref{fig:gra-ARPES}(c) and the EDC in Fig.~\ref{fig:gra-ARPES}(d) taken at $k_{\parallel}=$~\SI{0.81}{\per\angstrom}. The intrinsic nature of the graphene formed on this surface is, as far as we are aware, contrary to all calculations performed in the literature and in our own DFT calculations, which show an n-type doping for graphene \cite{Hu2013}. In pristine, undoped, and unbuckled graphene, the binding energy of the graphene $\pi\text{-band}$ at the $\overline{\text{M}}_{\text{G}}$-point should be $\approx$\SI{3}{\electronvolt}. However, in both the ARPES and the DFT modelling of graphene on the diamond C$(111)\text{-}(2\times1)$ surface presented here, the energy of the band at the $\overline {\text{M}}_{\text{G}}$-point is observed closer to $E_{\text{F}}$, i.e. $E_{\text{B}} =$~\SI{2.65}{\electronvolt}. This up-shift of the $\pi\text{-band}$ at the $\overline {\text{M}}_{\text{G}}\text{-point}$ (and by extension, a flattening of the dispersion along the {{$\overline{\text{K}}_{\text{G}}\text{-}\overline{\text{M}}_{\text{G}}\text{-}\overline{\text{K}}\textsubscript{\rlap{G}}'$}}~ edge of graphene's BZ) can potentially be attributed to an increased third-nearest atomic neighbour coupling due to either uniaxial pressure, shear, or strain forces, doping, or more likely in this situation, an increased graphene-substrate interaction \cite{Bena2011,Bostwick2007}.

Following optimisation of the atomic positions and the inclusion of VdW interactions in our DFT calculations, the graphene sheet moved to a stable distance of \SI{3.29}{\angstrom} away from the C$(111)\text{-}(2\times1)$ surface to avoid buckling as shown in Fig.~\ref{fig:struct-mod}(c), which is only slightly smaller than the graphite interlayer distance of \SI{3.35}{\angstrom}. This structural model contradicts previous experimental results that suggest an unreconstructed C$(111)\text{-}(1\times1)$ surface existing below the graphene formed at \SI{1250}{K} \cite{Ogawa2012}. We found that the inclusion of VdW interactions had little bearing on the electronic dispersions calculated using {\sc{Aimpro}} or {\sc{Vasp}}, as there were no discernible differences between them. However as shown in Fig.~\ref{fig:gra-ARPES}(f), the VdW interactions did play a role in determining the optimized graphene distance above the C$(111)\text{-}(2\times1)$ surface. The {\sc{Aimpro}} calculations showed no obvious minima in cohesion energy ($E_{\text{Cohesion}}$) as a function of distance away from the diamond surface.

To investigate if the graphene layer will remain intact following exposure of the substrate to atmosphere, and coinciding deconstruction of the diamond C$(111)\text{-}(2\times1)$ bonds, a study of the surface properties was conducted using XPS, UPS and LEED. XPS measurements of the C~1s core level acquired after reconstruction at \SI{930}{\degreeCelsius} and graphitization at \SI{1000}{\degreeCelsius} are presented in Fig.~\ref{fig:gra-XPS}(a) and (b) respectively. Corresponding widescan spectra are shown in Fig.~S4 in the supplementary information. A type Ia substrate is chosen here as the bulk C~1s core level for the type IIb sample has a very similar $E_{\text{B}}$ to that of graphene and the C$(111)\text{-}(2\times1)$ surface at $\approx$\SI{284}{\electronvolt}. The use of an insulating diamond, together with a supply of charge compensating secondary electrons from the unmonochromated twin-anode x-ray source, allows for a clear distinction to be made between the conductive $\text{sp}^2\text{-bonded}$ surface components and the bulk $\text{sp}^3\text{-bonded}$ diamond substrate.

\begin{figure}[!ht]
\centering
 \includegraphics[width=0.95\columnwidth]{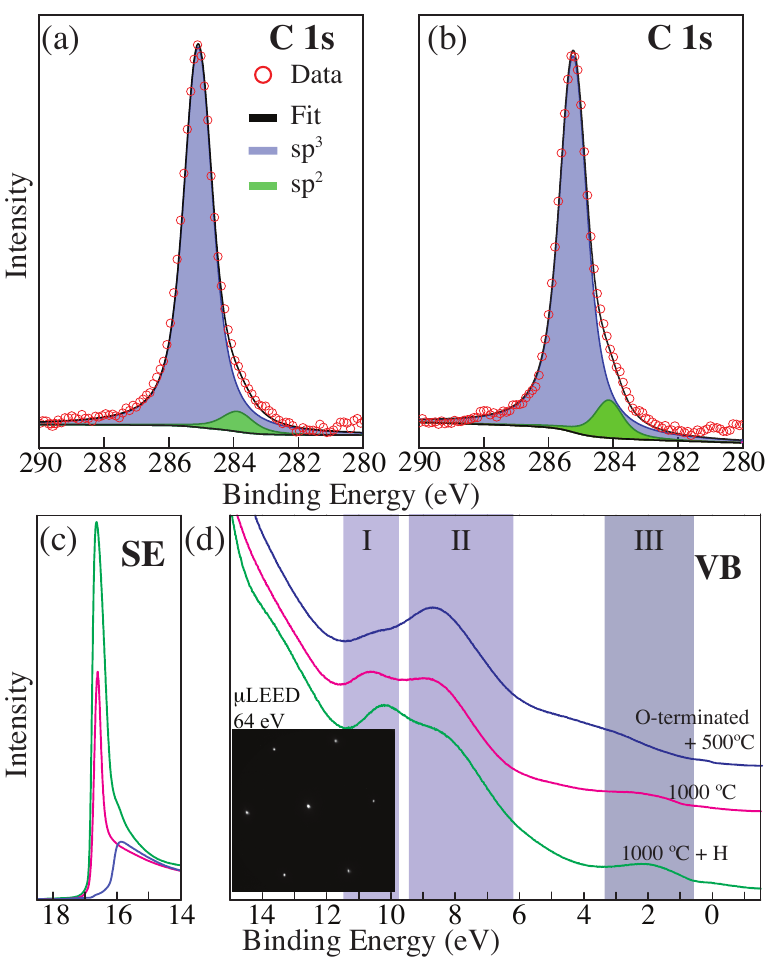}
 \caption{XPS and UPS measurements of the diamond surface following high temperature annealing and atomic hydrogen treatment. (a) and (b) Core level intensity for C~1s spectra collected at a pass energy of \SI{20}{\electronvolt} and $h\nu =$~\SI{1253.6}{\electronvolt} after \SI{950}{\degreeCelsius} and \SI{1000}{\degreeCelsius} respectively on the type Ia diamond substrate. (c) and (d) UPS spectra at various stages of preparation of the type IIb diamond substrate collected at a pass energy of \SI{5}{\electronvolt} and $h\nu =$~\SI{21.2}{\electronvolt}. (c) SE cut off and shape of the SE tail and (d) intensity of the integrated valence band. The regions labelled I and II are related to bulk $\sigma\text{-states}$ where region III relates to $\pi\text{-states}$. The inset panel in (d) shows a $\mu$LEED pattern of the surface following hydrogen termination of the graphene on the C$(111)\text{-}(2\times1)$ diamond system.}\label{fig:gra-XPS}
\end{figure}
Both the reconstructed diamond surface and the graphitized diamond surface show two components: the larger component at higher binding energy is naturally attributed to the $\text{sp}^3\text{-bonded}$ carbon of the diamond substrate shown in blue, and the smaller component at lower binding energy shown in green is consistent with $\text{sp}^2\text{-bonded}$ carbon species (i.e. at $E_{\text{B}} =$~\SI{284}{\electronvolt}). The $\text{sp}^2\text{-bonded}$ carbon component only emerges when the samples are subjected to temperatures greater than \SI{930}{\degreeCelsius}. After graphitization, the $\text{sp}^2$ component has a small degree of asymmetry, implying that there has been an increase in the density of states at $E_{\text{F}}$. This component doubled in intensity after the sample is heated to \SI{1000}{\degreeCelsius}, from \SI{5.4}{\percent} to \SI{10.6}{\percent} of the total C~1s area. This means that this component must account for two structures of $\text{sp}^2$ carbon: The C$(111)\text{-}(2\times1)$ reconstruction, and the graphene layer. If the intensity of the $\text{sp}^2$ peak in Fig.~\ref{fig:gra-XPS}(a) represents a single layer of $\text{sp}^2$ carbon due to the reconstruction, then it follows that a doubling of this intensity in Fig.~\ref{fig:gra-XPS}(b) represents a contribution from the graphene sheet, as two layers of reconstruction is not possible and sub-surface reordering has not been demonstrated within the literature.

Following annealing to \SI{1000}{\degreeCelsius} a new feature is also observed in the UPS spectra in region III of Fig.~\ref{fig:gra-XPS}(d) at $\approx$\SI{2}{\electronvolt}. This feature is attributed to the $\pi\text{-band}$s of the ($2\times1$) reconstruction and graphene as seen in the ARPES datasets. It is well known that hydrogen induces a deconstruction of the C$(111)\text{-}(2\times1)$ surface \cite{Kern1998,Kern1996,Kuettel1995}. However, following hydrogen treatment in our in-situ plasma system, the state at $\approx$\SI{2}{\electronvolt} is still evident, and appears to have increased in intensity, despite the fact that the secondary electron (SE) tail shown in Fig.~\ref{fig:gra-XPS}(c) has also dramatically increased in intensity, a typical characteristic of the H-terminated diamond (111) surface’s NEA \cite{Cui1998a}. $\mu$LEED measurements performed after similar in-situ treatment reveal that the surface is no longer reconstructed as the loss of second-order spots is evident when compared to Fig.~\ref{fig:2by1-surf}(f). A feature at $\approx$\SI{2}{\electronvolt} is also apparent in the density of states calculations for graphene adhered to a H-terminated diamond (111) surface by Zhao $et~al.$ \cite{Zhao2019}. We believe that this is strong evidence that the graphene grown in this manner survives the gentle (\SI{150}{\watt}) hydrogen plasma processing, which would allow for various diamond surface terminations to be achieved under the graphene, doping the graphene film via charge-transfer either n- or p-type depending on the terminating species \cite{Wan2020,Zhao2019,Ma2012,Hu2013}.

\section{Conclusions}

In summary, we have investigated the occupied electronic structure of the C$(111)\text{-}(2\times1)$ surface, experimentally using ARPES, and by DFT calculations using the {\sc{Aimpro}} and {\sc{Vasp}} codes. Our results show that the C$(111)\text{-}(2\times1)$ surface exhibits a metallic surface state that intersects $E_{\text{F}}$ at the $\overline {\text{K}}_{\text{I,II}}$-point. This observation is in good agreement with DFT modelling both in this work and the wider literature, in particular with Pandey’s original prediction of symmetric $\pi\text{-bonded}$ surface chains, but contrary to previous ARPES studies and dimerized surface chain models. The experimental observations are attributed to a combination of the exceptional surface quality of the diamond samples used in this study, their low miscut angle, and possibly to the higher resolution of the ARPES instrumentation used. In conjunction with relatively recent structural investigations that confirm the symmetric $\pi\text{-chain}$ model, we can now confidently say that a consistent and robust model of the C$(111)\text{-}(2\times1)$ surface exists that reconciles both its structural \emph{and} electronic properties.

New states observed in the ARPES Fermi surface measurements after annealing the C$(111)\text{-}(2\times1)$ surface to \SI{1000}{\degreeCelsius} in vacuo, are attributed to the liberation of the surface atoms to a monolayer of graphene that coexists above a newly reconstructed C$(111)\text{-}(2\times1)$ surface. The graphene appears to be undoped with its $\overline {\text{K}}_{\text{G}}$-point intersecting $E_{\text{F}}$ within experimental resolution. This contrasts with all calculations suggesting there should be a n-type doping of at least \SI{100}{\milli\electronvolt} for graphene on the C$(111)\text{-}(2\times1)$ surface. An increased graphene-substrate interaction is also observed, evidenced by a flattening of the $\pi\text{-band}$ dispersion between the $\overline {\text{K}}_{\text{G}}$ and $\overline {\text{M}}_{\text{G}}$ points. Deconstruction of the diamond surface to a H-terminated C$(111)\text{-}(1\times1)$ structure after gentle plasma treatment does not affect the graphene, providing the option for device manufacturers to tailor the doping of the graphene sheet by altering the termination of the diamond surface bonds beneath the graphene.

\begin{acknowledgments}
We would like to acknowledge the financial support by: the EPSRC CDT in Diamond Science and Technology; The European Regional Development Fund through the Solar Photovoltaic Academic Research Consortium (SPARC II) operated by the Welsh Government, the UK research council EPSRC grant \#EP/G068216/1; The Research Council of Norway through the Centres of Excellence scheme, project \#262633 “QuSpin” and FriPro ToppForsk-program \#251131 FUNDAMeNT; the National Infrastructure for High Performance Computing and Data Storage in Norway UNINETT Sigma2; and the National Measurement System programme of the UK Department of Business, Energy and Industrial Strategy. We would also like to acknowledge the technical support from Dr Marco Bianchi and insightful discussions held with Prof. Philip Hofmann at the ISA synchrotron in Aarhus, and Dr Federico Mazzola for the many fruitful discussions and helping implement the code used to simulate the ARPES results from DFT calculations. MEB would like to acknowledge an ETH Zurich Postdoctoral Fellowship program.
\end{acknowledgments}

\bibliography{Main.bib}

\end{document}